# Precession-free domain wall dynamics in compensated ferrimagnets

E. Haltz[1], J. Sampaio[1], S. Krishnia[1], L. Berges[1], R. Weil[1] and A. Mougin[1, *]

[1] Laboratoire de Physique des Solides, CNRS, Univ. Paris-Sud, Université Paris-Saclay Bâtiment 510, 91405 Orsay
* : alexandra.mougin@u-psud.fr

**One fundamental obstacle to efficient ferromagnetic spintronics is magnetic precession, which intrinsically limits the dynamics of magnetic textures, We demonstrate that the domain wall precession fully vanishes with a record mobility when the net angular momentum is compensated ($T_{AC}$) in DWs driven by spin-orbit torque in a ferrimagnetic GdFeCo/Pt track. We use transverse in-plane fields to reveal the internal structure of DWs and provide a robust and parameter-free measurement of $T_{AC}$. Our results highlight the mechanism of faster and more efficient dynamics in materials with multiple spin lattices and reduced net angular momentum, promising for high-speed, low-power spintronics applications.**

In magnetic materials, the exchange interaction aligns the magnetic moments producing ferromagnetic or antiferromagnetic orders. Even if ferromagnets have numerous applications in spintronics, magnetic precession and stray fields limits the development of higher-density and faster devices. Antiferromagnetic order leads to faster dynamics and insusceptibility to spurious fields, and is emerging as a new paradigm for spintronics [1]. However, antiferromagnets are hard to probe due to their tiny magnetization, and therefore have been rarely studied or used in applications. Rare Earth-Transition Metal (RETM) ferrimagnetic alloys allow to benefit from antiferromagnetic-like dynamics and ferromagnetic-like spin transport. Indeed, they have two antiferromagnetically-coupled spin sublattices, corresponding roughly to the RE and TM moments, and their spin transport is carried only by the TM sub-lattice. RETM thin films exhibit perpendicular magnetic anisotropy, are conductors, and present large spin transport polarization and spin torques, even when integrated in complex stacks [2–8]. Furthermore, its magnetic properties can be controlled by changing either its composition or temperature, as described by the mean-field theory [9] in Fig 1a. For a given composition, there may be two characteristic temperatures: the angular momentum compensation temperature ($T_{AC}$) for which the net angular momentum is zero, and the magnetic compensation temperature ($T_{MC}$), for which the net magnetization ($M_S$) is zero (Fig. 1a). Interestingly, due to the two different Landé factors, these two temperatures are different. At $T_{MC}$ the magnetostatic response diverges (as observed in the coercivity, anisotropy field, …). In contrast, at $T_{AC}$ the dynamics is affected, changing the precession rate and direction (as this letter reports). Although these effects are challenging to evidence, the singular and promising behaviour of RETM at $T_{AC}$ was observed in current-induced switching [10], magnetic resonance [11], and time-resolved laser pump-probe measurements [11,12]. Very recent reports have revealed high domain wall (DW) velocities close to $T_{AC}$ [4,5,13–15]. However, the strong sensitivity of DW propagation to Joule heating and pinning [3,16] has impeded a precise determination of $T_{AC}$ and a better understanding of the DW dynamics in compensated ferrimagnets.

DWs driven by spin-orbit torque (SOT) have been observed in thin magnetic films with a heavy-metal adjacent layer, like Pt, which induces three main interfacial effects: perpendicular anisotropy, Dzyaloshinskii-Moriya exchange interaction (DMI), and vertical spin current generated by the spin Hall effect (SHE) (Fig. 1b). Such systems present chiral Néel DWs, which is the configuration for which the SOT DW driving is effective (Fig 1b) [17].

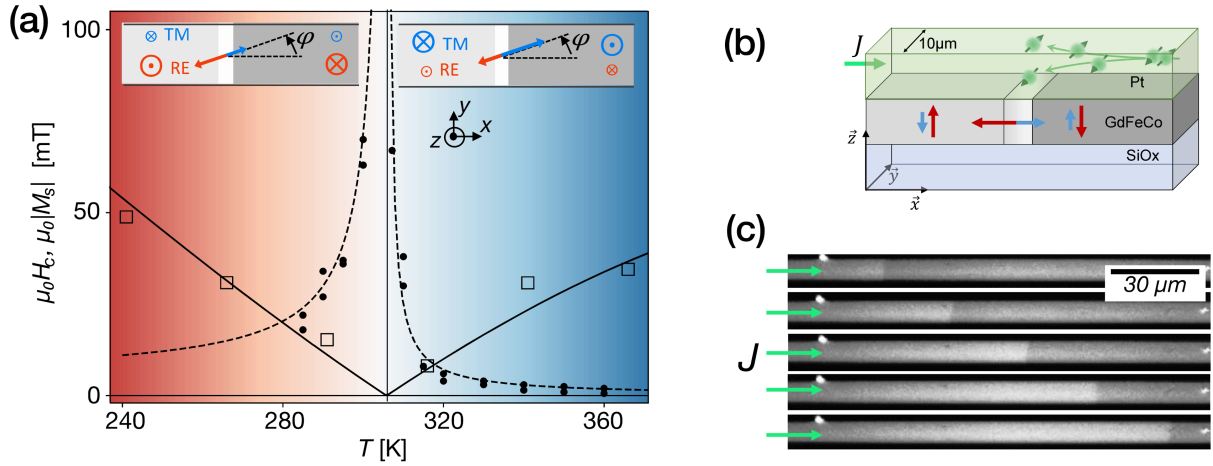

**Figure 1 GdFeCo/Pt sample properties and SOT-driven DW propagation in tracks.** (**a**) Measured net magnetisation $M_S$ (squares) of the virgin film and coercivity $H_C$ (dots) of the patterned track versus temperature $T$. The $M_S$ points were shifted by -25 K to account for depletion of Gd during patterning [2]. The continuous line is the result of the mean-field model (see suppl.). The background highlights the dominant sublattice. Inset: sketch of the magnetisation of the two sublattices (in red and blue) below and above $T_{MC}$. The size of the arrows represents their relative magnitude. The grayscale corresponds to the domain Kerr contrast while the DW is depicted in white. The angle of the DW magnetisation is given by $\varphi$. (**b**) Sketch of the track containing a SOT-driven DW. (**c**) Kerr images of a DW driven by SOT, with 25 ns long pulses of $J$=300 GA/m².

To investigate SOT-driven DW dynamics in RETM, a 10 µm-track of amorphous GdFeCo(5nm)/Pt(7nm) with perpendicular magnetic anisotropy (Fig. 1a and 1b) was fabricated. The film of amorphous GdFeCo (5nm) capped with Pt (7nm) was deposited by electron beam co-evaporation in ultrahigh vacuum on thermally-oxidised Si substrates. Details of the film growth and characterisation can be found in [2]. The tracks were patterned by e-beam lithography and hard-mask ion-beam etching. Transport measurements of the extraordinary Hall effect (EHE) versus field were made on 5 µm wide crosses at different temperatures in a commercial QD PPMS. $M_S(T)$ was measured by SQUID magnetometry. The $T_{MC}$ of the material, 308 K, was determined by measuring the coercivity, magnetization and sign of the magneto-optical Kerr effect (MOKE) and EHE loops (Fig 1a). The magnitude of DMI equivalent field $H_{DMI}$ was determined by analysing the SOT-driven DW velocity with a field collinear to the current as in [18]. The magnitude of the SOT equivalent field $H_{DL}$ was determined with the harmonic voltage method [19,20]. Kerr microscopy experiments were performed using an adapted commercial Schafer Kerr microscope, with a controlled temperature sample holder (at temperature $T_{SP}$.) 25 ns pulses of current density $J$ were applied in the track containing an up-down DW. After each pulse, a Kerr image was recorded (Fig. 1c). The DWs move against the electron flow, which rules out any significant spin-transfer torque [3] and is compatible with SOT-driving of chiral Néel DWs with the same relative sign of DMI and SHE as found in ferromagnetic Pt/Co [17,21]. The linearity of the DW displacement with the pulse number and duration (see suppl.) allows the robust determination of the propagation velocity $v$. High DW velocities (>700m/s; Fig 2a) are observed for low $J$ (~600GA/m²), as previously reported in similar alloys [4,5]. Two significant differences are observed between the measured $v(J)$ and the theoretical predictions shown in the inset of Fig. 2a. These calculations were done using the 1D model described in [17] in the steady-state regime ($\dot{\varphi} = 0$), extended to include external magnetic fields and neglecting the in-plane demagnetisation field (see Supplemental Material [22]).

Firstly, DW motion occurs only above a threshold $J$ of few tens of GA/m² that we attribute to DW pinning at defects, supported by the fact that the threshold $J$ decreases when $T_{SP}$ increases. Similar behaviour is often found in current-driven DWs, both in ferromagnets [23,24] and in ferrimagnets [4,5]. In our wire, the threshold current is about 60 GA/m², a few times lower than in previous studies [4,5,24]. Secondly, for a given $T_{SP}$, the velocity exhibits a non-monotonous behaviour (Fig 2a), while it is expected that the velocity is always increasing [17,23]. The $v(J)$ measured at fixed $T_{SP}$ (Fig. 2a) can be understood by considering the theoretical $v(J)$ curves for different $T$ (Fig. 2a inset). Each measurement at fixed $T_{SP}$ corresponds to a point in a curve $v(J)$ of progressively higher $T$ as $J$ increases, which produces a peak. To avoid this, we plot the mobility $\mu=v/J$ versus $T_{SP}$ for different $J$ (Fig.2b). In this representation, the Joule heating induces a simple horizontal shift between curves. For a given $J$, we observe a peak of mobility (marked by a star), up to 1.2 (m/s)/(GA/m²) (ten times higher than in previous reports [4,5,24]). Fig.2c shows all measured mobilities in a $(J,T_{sp})$ colour-plot, and for each $J$ the maximum $\mu$ is marked with a star. The coordinates of the maxima $\mu$ in Fig. 2c follow $T_{SP} - T \propto J^2$ (solid line in Fig. 2c), which suggests that they all occur at a single track temperature $T$.

Models predict that the SOT-driven DW velocity follows

$$v/J \propto \cos(\varphi) \quad \text{(eq. 1)}$$

where $\varphi$ is the angle of the internal DW magnetisation [17]. In ferromagnets, $\varphi$ is determined by the balance between DMI, which favours the Néel configuration ($\varphi=0$), and the precession induced by SOT, which increases $|\varphi|$. In ferrimagnets, the precession depends on temperature and is expected to vanish at $T_{AC}$. Therefore, the observed mobility peak corresponds to minimal $|\varphi|$, and it can be deduced that the temperature of the maxima is $T_{AC}$ (342 K according to the fit in Fig. 2c), as previously done in refs. [4,5]. However, DW mobility is affected by all other forces on the DW, such as Oersted fields or thermal depinning, which can shift the maxima. Indeed, Hirata and colleagues [16] reported that the temperature of the mobility peak is affected by pinning in similar experiments measuring field-driven DW motion. To overcome this, we propose a new method based on the application of a transverse field $H_Y$ that, in addition, reveals the internal magnetic dynamics of the DW across both compensation points. Simultaneously, it determines the Joule heating amplitude.

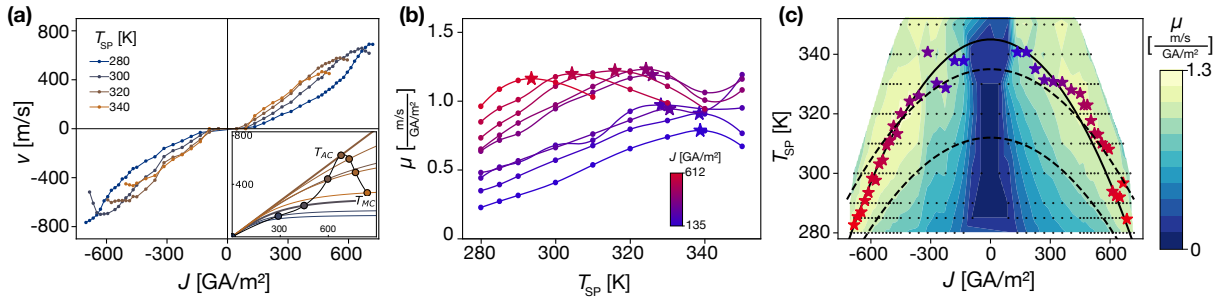

**Figure 2 | DW mobility versus current and temperature in a GdFeCo/Pt track. (a)** DW velocity versus current density, $v(J)$, for different holder temperatures $T_{SP}$. **Inset:** theoretical $v(J)$ at constant track temperature $T$ (blue to orange lines), and sketch of a $v(J)$ at constant $T_{SP}$ considering Joule heating (coloured dots and black line). **(b)** DW mobility $\mu=v/J$ versus $T_{SP}$ for some values of $J$. Maximum mobilities are marked by stars in (b) and (c). **(c)** $(J,T_{sp})$ colour-plot of all measured mobilities. The solid line corresponds to the quadratic fit of the maximum mobilities, with $T_{SP}$ = 342 K - 0.00013$J^2$. The dashed lines are the fits of Fig. 3c.

We measured the DW velocity $v$ versus $T_{SP}$ with an applied in-plane field $H_Y$ perpendicular to the current flow. Fig. 3a shows $v(T_{SP}, H_Y)$ for $\mu_0 H_Y = 0$ and $\pm 90$ mT ($J$=360 GA/m²; see suppl. mat. for other values of $J$). With $H_Y$, as with $H_Y = 0$, the DW moves along the current direction but, for some ranges of $T_{SP}$, the DW is faster with positive $H_Y$ while in other ranges it is faster with negative $H_Y$. Two crossing points, $T_{SP,1}$ and $T_{SP,2}$, are observed where $v(T_{SP,i}, +H_Y) = v(T_{SP,i}, -H_Y)$. They are more readily distinguished by plotting $\Delta v(T_{SP}) \equiv v(T_{SP}, +H_Y) - v(T_{SP}, -H_Y)$, shown in Fig. 3b ($T_{SP,1}$ = 300K and $T_{SP,2}$ = 328K). Note that $T_{SP,1}$ and $T_{SP,2}$ are lower for higher $J$ (Fig. 3c) but the difference $T_{SP,2} - T_{SP,1}$ seems independent of $J$ (see suppl.). These observations hold for both current polarities, and for different magnitudes of $H_Y$ (see suppl.).

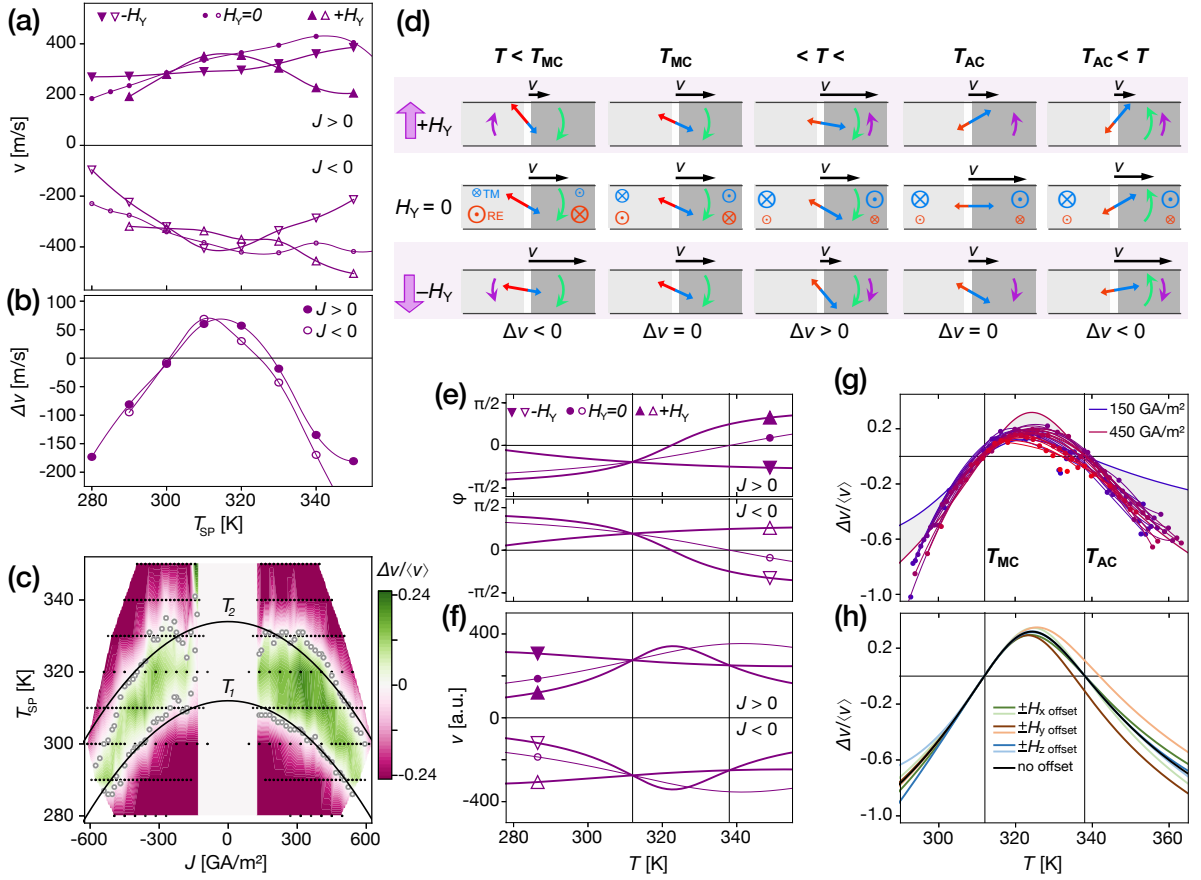

**Figure 3 | SOT-driven DW under $H_Y$: determination of the internal DW dynamics, $T_{MC}$ and $T_{AC}$.** (a) Measured DW velocity $v$ versus sample holder temperature $T_{SP}$ with $H_Y = \pm 90$ mT (▲,▼) or 0 mT (●), for $J=\pm 360$ GA/m². (b) Velocity difference $\Delta v(T_{SP}) \equiv v(J,+H_Y)-v(J,-H_Y)$ for the same $H_Y$ and J of (a). (c) Colour-plot of $\Delta v(J,T_{SP})$. Black points correspond to measurements. White points correspond to crossing points where $\Delta v=0$ ($T_{SP,1}$ and $T_{SP,2}$). The black lines are the parabolic fits of $T_{SP,1}$ and $T_{SP,2}$ ($T_{SP,i} = T_i - a J^2$, $a = 0.00008$, $T_2 = 312$ K, $T_2 = 334$ K). (d) Diagram of the sublattice orientations in a SOT-driven DW under $H_Y$ across compensation points. The red and blue arrows correspond to RE and TM, respectively. The purple curved arrows represent the torque due to $H_Y$, and the green ones due to SOT. (e, f) Theoretical DW angle $\varphi$ and DW velocity versus temperature. (g) Normalised velocity difference $\Delta v/\langle v \rangle$ versus track temperature T taken with different substrate temperature $T_{SP}$ ($\langle v \rangle = (v(J,+H_Y) + v(J,-H_Y))/2$). The temperature T was determined with the Joule heating law of (c). The envelope corresponds to the calculated $\Delta v/\langle v \rangle$ for $150 < J < 450$ GA/m². (h) Theoretical $\Delta v/\langle v \rangle$ with an offset bias field of 10 mT along x, y or z ($|H_Y|$ = 90 mT and $J$=360 GA/m²).

In Fig. 3c, all measured $\Delta v$ are shown in a colour-plot as a function of $T_{SP}$ and $J$. Three regions can be observed with, successively, $\Delta v<0$, $\Delta v>0$ and $\Delta v<0$, separated by the two sets of crossing points $T_{SP,1}$ and $T_{SP,2}$. Both follow a Joule heating parabolic relation, $T_{SP,i}(J) = T_i - a J^2$, which shows that the crossing points occur at the same track temperatures $T_i$ independently of $J$. The fits of $T_{SP,1}(J)$ and $T_{SP,2}(J)$ (parallel parabolic lines in Fig 3c) are independent and give the same heating parameter $a$ (within 1%), showing that $T_{SP,2} - T_{SP,1}$ is indeed constant for all $J$ (see suppl. mat.). The corresponding track temperatures were determined: $T_1$= 312 K and $T_2$=334 K.

To understand the effect of $H_Y$ on SOT-driven DWs in ferrimagnets, we first consider the well-understood ferromagnetic case. The transverse field $H_Y$ changes the velocity by affecting the angle of the DW magnetisation $\varphi$ [18]. The balance between the DMI (that stabilises the Néel configuration, $\varphi = 0$), the torques induced by current (SOT) and the external field (through Zeeman interaction) gives:

$$\varphi = \arctan\left(\frac{\Delta}{D}\left(\frac{\hbar}{2et}\frac{\theta_{SHE}}{\alpha}J + \mu_0 M_s H_y\right)\right) \quad \text{(eq. 2)}$$

where $\Delta$ is the DW width parameter, $D$ is the DMI parameter, $\alpha$ is the Gilbert damping parameter, $\hbar$ is the reduced Planck constant, $e$ is the fundamental charge, $\theta_{SHE}$ is the spin Hall angle of the Pt layer, and $t$ is the magnetic film thickness. As $v \propto \cos\varphi$ (eq. 1), $\Delta v$ clearly shows whether $H_Y$ rotates $\varphi$ closer to or farther from the Néel configuration. A positive $\Delta v$ means that $J$ and $+H_Y$ have opposite contributions to $\varphi$, and that $+H_Y$ brings the DW closer to the Néel configuration ($\varphi \to 0$, eq. 2).

In the RETM ferrimagnetic case, the field acts on both RE and TM sub-lattices, whereas spin current interact only with the TM sub-lattice (ref. [2] and references therein). The DW velocity can still be described with the same model using the $\varphi$ of the TM sub-lattice and effective parameters of the alloy: $M_S \leftarrow (M_{TM} - M_{RE})$ and $\alpha \leftarrow \alpha_{eff}$ [11,12,25]. The sign of these parameters change with the temperature, $M_S$ at $T_{MC}$ and $\alpha_{eff}$ at $T_{AC}$ (see suppl.). The sketch in Fig. 3d shows the effects of $J$ (green arrows) and $H_Y$ (purple arrows) on $\varphi$ for different temperatures. Knowing beforehand the sign of $M_S$ and measuring $\Delta v$, the sign of the effect of $J$ can be determined. At $T<T_{MC}$, the RE sublattice is dominant ($M_{RE} > M_{TM}$) and $M_S$ is negative. $+H_Y$ rotates $\varphi$ clockwise (CW). A $\Delta v<0$ means that the current contribution to $\varphi$ is also CW (1st column of Fig. 3d). Above $T_{MC}$, $M_S$ becomes positive and the effect of external fields is reversed. Now, a positive $\Delta v$ means that the current contribution to $\varphi$ is counterclockwise (CCW), whereas a negative $\Delta v$ means that the current contribution is CW. At $T=T_{MC}$, $M_S$ is 0 and $H_Y$ affects neither $\varphi$ nor $v$ (2nd column of Fig. 3d) and $\Delta v=0$. The remarkable dynamic properties of ferrimagnets drastically change at $T=T_{AC}$; in particular, $\alpha_{eff}$ is expected to diverge and change sign [11,12,25]. Consequently, from eq. 2, the current contribution to $\varphi$ should change sign across $T_{AC}$. Furthermore, at $T=T_{AC}$, the current effect on $\varphi$ should vanish, *i.e.* the DW should remain Néel (without field). A positive and negative $H_Y$ will induce a $\varphi$ of opposite sign but of equal amplitude, decreasing equally the velocity, and thus $\Delta v = 0$ (4th columns of Fig. 3d). In a ferrimagnet, therefore, $\Delta v$ changes sign at $T_{MC}$ and $T_{AC}$. In the measurements, the two crossing points $T_1$ and $T_2$ correspond to $T_{MC}$ or $T_{AC}$, separating the three types of dynamics described in Fig. 3d. In particular, at $T_1$, $v(\pm H_Y) = v(H_Y=0)$ (Fig. 3a), as it is expected at $T_{MC}$. At the second crossing point $T_2$: $v(\pm H_Y) < v(H_Y=0)$ as it is expected at $T_{AC}$ (Fig. 3a). Note that $T_{MC}$ and $T_{AC}$ are consistent with the previous measurements of $H_C(T)$ and $\mu(J)$ in Fig. 1 ($T_{MC}$ = 308 K) and Fig. 2 ($T_{AC}$ = 342 K).

All the normalised $\Delta v/\langle v \rangle$ can be superimposed on the same graph versus $T$ in Fig. 3g using the obtained Joule heating law. The envelope corresponds to the calculated $\Delta v/\langle v \rangle$ for $J$ between 150 and 450 GA/m² using eq. 1 and 2, with an excellent agreement. Unlike the method of the peak DW SOT mobility (Fig. 2), this determination of $T_{AC}$ is based on the intrinsic DW dynamics and so it is unaffected by pinning. At $T_{AC}$, the DW is sensitive to external fields since $M_s$ is not zero. Spurious external field has impact only along the y direction, as confirmed in the calculated $\Delta v/<v>$ with an offset external field (Fig. 3h), which explains the data dispersion at this point at $T_{AC}$.

This approach determines precisely the sense of the DW precession using a very large difference of DW velocities ($\Delta v \approx 100$ m/s) that gives $T_{MC}$ and $T_{AC}$ without requiring any material parameters. Figs. 3e,f show analytical calculations of the DW angle $\varphi$ and related velocity $v$ as a function of $T$ and $H_Y$. All material parameters were taken from measurements except for $\alpha(T)$ (see Supplemental Materials [22]). The excellent agreement between the velocity measurements in Fig. 3a and the calculations in Fig. 3f allow us to evaluate the DW internal angle *φ(T)*. It is possible to reconstruct the internal state of the moving DW without field. At $T = T_{AC}$, the SOT-driven DW remains Neel (*φ*=0) and the propagation is precession-free which shows that the effective damping $α_{eff}$ diverges (Eq. 2). Across $T_{AC}$, the SOT-induced angle changes sign, evidencing that $α_{eff}$ also changes sign. The vanishing magnetic precession enables the observed record DW mobility in a compensated RETM ferrimagnet. This opens new perspectives for fast and energy-efficient spintronics using any angular-momentum-compensated multi-lattice material.

## Acknowledgements

We are very thankful to S. Rohart and A. Thiaville for fruitful discussions, and to R. Mattana for the SQUID measurements of $M_S$. S. K. and E. H. acknowledge public grant overseen by the ANR as part of the "Investissements d'Avenir" programme (Labex NanoSaclay, reference: ANR-10-LABX-0035) for the FEMINIST project and travelling grants. The transport measurements were supported by Université Paris-Sud Grant MRM PMP.

# Supplementary materials to
# "Precession-free domain wall dynamics in compensated ferrimagnets"

E. Haltz[1], J. Sampaio[1], S. Krishnia[1], L. Berges[1], R. Weil[1] and A. Mougin[1]

[1] Laboratoire de Physique des Solides, CNRS, Univ. Paris-Sud, Université Paris-Saclay Bâtiment 510, 91405 Orsay

*This supplementary contains additional data on the raw sample properties, mean field calculations, the extended equations used for analytical modelling with associated DW mobilities and a few additional experimental observations on DW propagation.*

*Magnetization measurements:*

The raw GdFeCo/Pt sample properties have been measured over a larger temperature range than the one shown in the main text. Results are shown in Fig. S1.

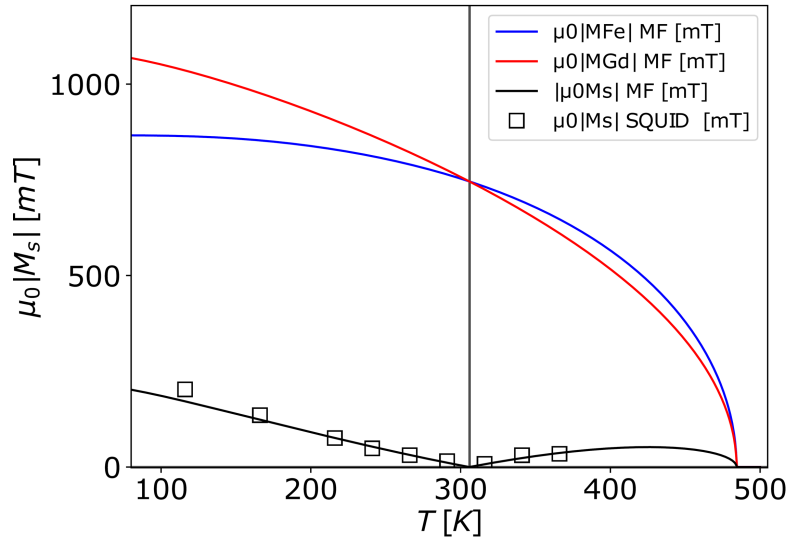

**Fig. S1**: Over a large temperature T range, mean-field (MF) calculations and measurements (SQUID) of |Ms|.

*Analytical model of DW velocity under SOT and field.*

In the main text, Eqs. 1 & 2 and the theoretical plots in Figs. 2 & 3, were done using the 1D model described in [1] in the steady-state regime ($\dot{\varphi} = 0$), extended to include external magnetic fields and neglecting the in-plane demagnetisation field:

$$\begin{cases} \dfrac{\alpha v}{\Delta} = \gamma_0 \left(H_Z + \dfrac{\pi}{2} H_{SHE} \cos\varphi\right) \\ \dfrac{v}{\Delta} = \gamma_0 \dfrac{\pi}{2}\left((H_{DMI} + H_X)\sin\varphi + H_Y \cos\varphi\right) \end{cases} \Leftrightarrow \begin{cases} v = \dfrac{\gamma_0 \Delta}{\alpha}\left(H_Z + \dfrac{\pi}{2} H_{SHE}\cos\varphi\right) \\ H_Z + (H_{SHE} - \alpha H_Y)\dfrac{\pi}{2}\cos\varphi = \alpha \dfrac{\pi}{2}(H_{DMI} + H_X)\sin\varphi \end{cases}$$

with $H_{SHE} = \dfrac{\hbar}{2e}\dfrac{\theta_{SHE}}{\mu_0 M_S t}J$, $H_{DMI} = \dfrac{D}{\Delta \mu_0 M_S}$.

In the absence of $H_Z$, this yields:

$$v = \frac{\gamma_0 \Delta}{\alpha} \frac{\pi}{2} H_{SHE} \cos \varphi, \qquad \varphi = \arctan\left(\frac{H_{SHE}/\alpha + H_Y}{H_{DMI} + H_X}\right)$$

with $H_Z$ smaller than the Walker field, this yields:

$$v = \frac{\gamma_0 \Delta}{\alpha} \frac{((H_{DMI}+H_X)^2 + H_Y^2)H_Z \alpha^2 + H_{SHE}\left(H_Y H_Z \alpha + \alpha(H_{DMI}+H_X)\sqrt{\frac{\pi^2}{4}A - H_Z^2}\right)}{A},$$

with $A = H_{SHE}^2 + 2\alpha H_{SHE} H_Y + \alpha^2((H_{DMI} + H_X)^2 + H_Y^2)$. These equations can be used for ferrimagnets using the effective parameters [2–4] as described above. The calculated plots in Fig. 3 are obtained using a constant ratio $D/\Delta$ obtained from the determination of $H_{DMI}$ ($\frac{D}{\Delta} = \mu_0 M_S(T) H_{DMI}(T) = 33\ kJ/m^3$), and the SOT factor $\frac{\hbar}{2e}\frac{\theta_{SHE}}{t}$ from the determination of $H_{DL}$ ($\frac{\hbar}{2e}\frac{\theta_{SHE}}{t} = \mu_0 M_S(T) H_{DL}(T)/J = 1.75\ \frac{kJ/m^3}{100\ GA/m^2}$). The only parameter that is not experimentally determined, $\alpha(T)$, is approximated by an inverse linear law $\alpha(T) = \frac{12.0\ K}{T - T_{AC}}$, chosen to best reproduce the shape of the experimental curves (see Fig. 3a and f).

*Sample parameters (measured and calculated):*

Several sample parameters are shown in Fig. S2 versus temperature. The net magnetization and the gyromagnetic ratio $\gamma_{eff}$ change sign at $T_{MC}$ whereas the angular momentum and the effective damping $\alpha_{eff}$ change sign at $T_{AC}$. The static parameters ($H_{DMI}$ and $H_{SHE}$) change sign and diverge at $T_{MC}$ whereas the dynamic ones ($\gamma_{0eff}$ and $\alpha_{eff}$) change sign and diverge at $T_{AC}$. SOT driven domain wall mobility shows a peak at $T_{AC}$.

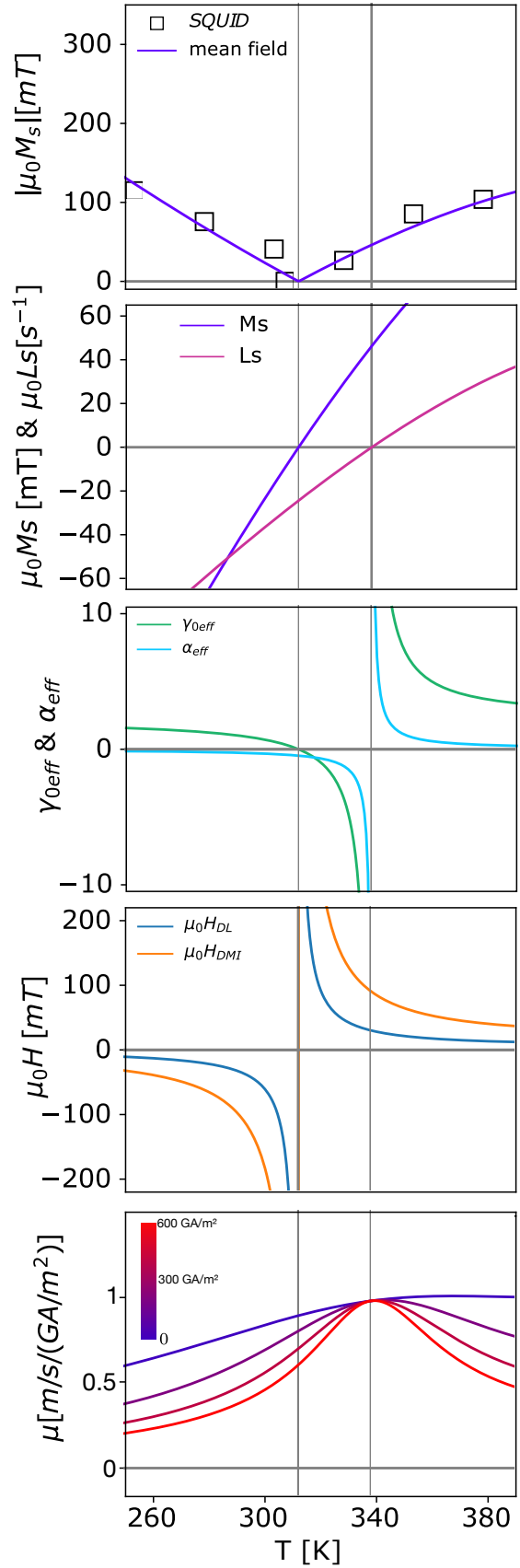

**Fig S2:** From top to bottom, over a temperature range close to that of experiments: Mean-field calculations and measurements of |Ms|; net magnetization and angular momentum; effective gyromagnetic ratio $\gamma_{0\,eff}$ ($10^5$ m/As) and damping $\alpha_{eff}$; $H_{DMI}$ and $H_{DL}$ (for J=360GA/m²); calculated mobilities for 3 current densities.

*Velocities measurements versus pulse duration:*

The linearity of the DW displacement with the number of pulses and with the pulse duration allows the determination of the propagation velocity *v*. As shown in Fig. S3, the velocity does not depend on the pulse duration, whatever the current density.

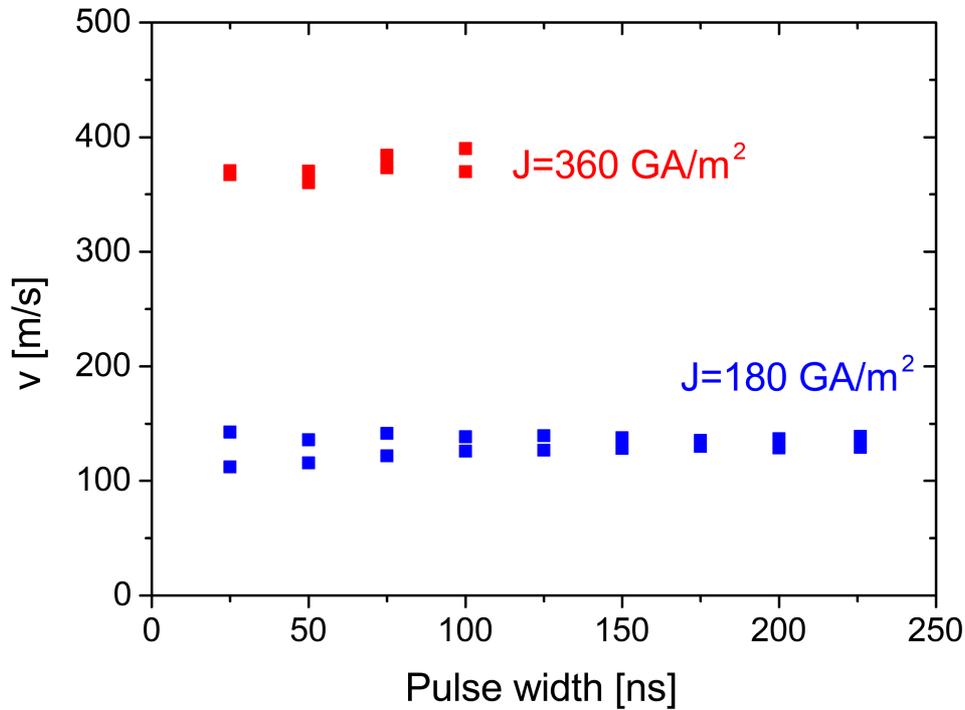

**Fig. S3:** DW velocity versus pulse duration for two current densities indicated in the figure.

*Velocities measurements with a field $H_Y$ versus $T_{SP}$ for different current densities or applied fields:*

We measured the DW velocity *v* versus $T_{SP}$ with an applied in-plane field $H_Y$ perpendicular to the current flow. In the main text, velocities are shown for $J$=360 GA/m$^2$; and here Fig. S4 shows other values of *J*.

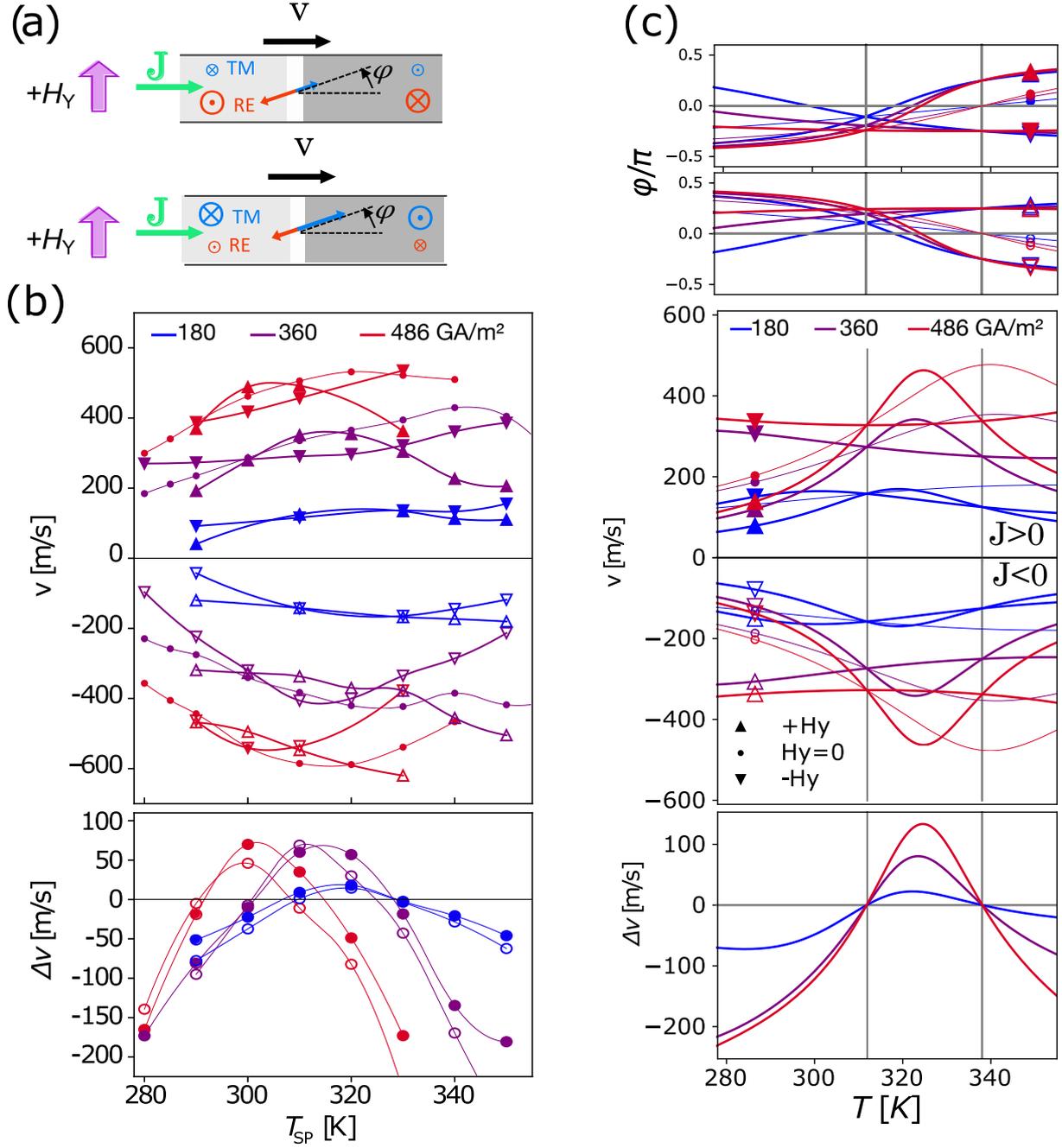

**Fig. S4: (a)** Sketch of a SOT-driven DW under $H_Y$. The red and blue arrows correspond to RE and TM, respectively below and above $T_{MC}$. The size of the arrows represents their relative magnitude. The grayscale corresponds to the domain Kerr contrast while the DW is depicted in white. The angle of the DW magnetisation is given by $\varphi$. The purple arrows represent $H_Y$, and the green ones the current. **(b)** DW velocity $v$ versus sample holder temperature $T_{SP}$ with $H_Y = \pm 90$ mT (▲,▼) or 0 mT (●), for $\pm J$ (top and bottom parts) and velocity difference $\Delta v(T_{SP}) \equiv v(J,+H_Y)-v(J,-H_Y)$ for the same $H_Y$ and $J$. **(c)** Calculated DW angle $\varphi$, DW velocity $v \propto \cos\varphi$ using the equations in the text and velocity difference $\Delta v$ versus track temperature $T$.

When an in-plane field is applied in DW propagation experiments, two crossing points, $T_{SP,1}$ and $T_{SP,2}$, are observed where $v(T_{SP,i}, +H_Y) = v(T_{SP,i}, -H_Y)$. Fig S5 shows the difference measured between the temperatures of these two crossing-points. Experiments were done with several values of the in-plane fields and results are shown in Fig. S6. The crossing points do not change.

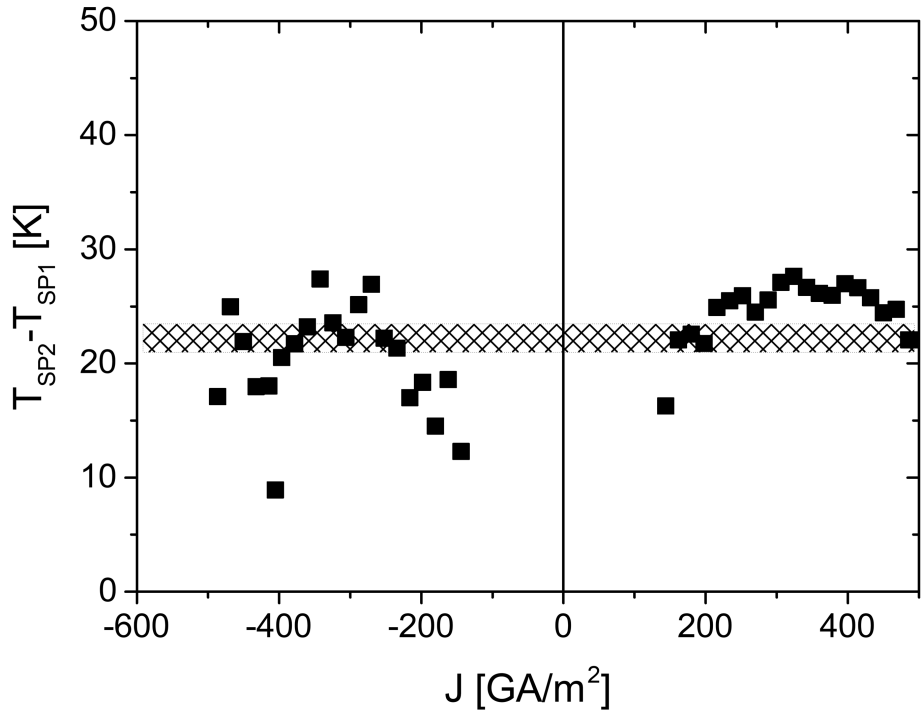

**FIG S5:** $T_{SP,2} - T_{SP,1}$ vs $J$ for SOT driven DW with an in-plane field.

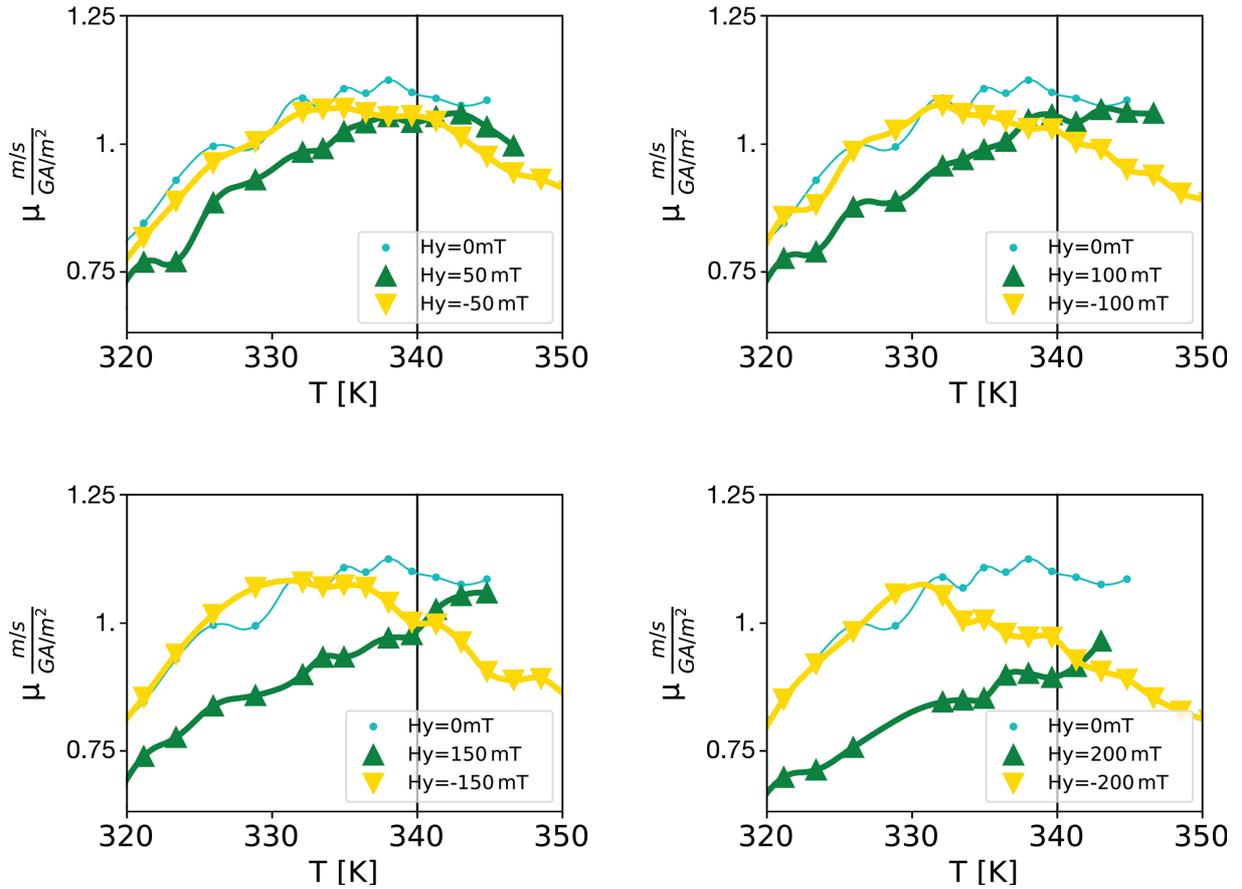

**FIG S6:** DW mobility (μ) versus $T$ for different Hy at $T_{SP}$=315 K and different $J$.